\documentstyle[12pt,epsfig]{article}
\begin{document}

\setlength{\topmargin}{-0cm}
\setlength{\headsep}{1.6cm}
\setlength{\evensidemargin}{.7cm}
\setlength{\oddsidemargin}{.7cm}
\setlength{\textheight}{22.5cm}
\setlength{\textwidth}{15.2cm}
\newcommand{\be}{\begin{equation}}
\newcommand{\ee}{\end{equation}}
\newcommand{\ba}{\begin{eqnarray}}
\newcommand{\ea}{\end{eqnarray}}

\title{Market Fluctuations II: multiplicative and percolation models, 
size effects and predictions}
\thispagestyle{empty}

\author{D. Sornette$^{1,2}$, D. Stauffer$^3$ and H. Takayasu$^4$ \\
$^1$ Institute of Geophysics and
Planetary Physics \\and Department of Earth and Space Science
\\ University of California, Los Angeles, California 90095\\
$^2$ Laboratoire de Physique de la Mati\`{e}re Condens\'{e}e\\ CNRS UMR6622 and
Universit\'{e} de Nice-Sophia Antipolis\\ B.P. 71, Parc
Valrose, 06108 Nice Cedex 2, France \\
$^3$ Institute for Theoretical Physics\\ Cologne University
50923 K\"oln, Germany\\
$^4$ Sony CSL, 3-14-13 Higashigotanda, Shinagawa-ku\\ Tokyo 141-0022, Japan
}

\maketitle

\begin{abstract}

We present a set of models of 
 the main stylized facts of market price fluctuations.
These models comprise dynamical evolution with threshold dynamics
and Langevin price equation with multiplicative noise, percolation models
to describe the interaction between traders and hierarchical cascade models
to unravel the possible correlation accross time scales, including 
the log-periodic signatures associated to financial crashes. The main
empirical knowledge is summarized and some key empirical tests are presented.

\end{abstract}

\newpage

\pagenumbering{arabic}

\section{Stylized facts of financial time series}

\begin{figure}
\begin{center}
\epsfig{file=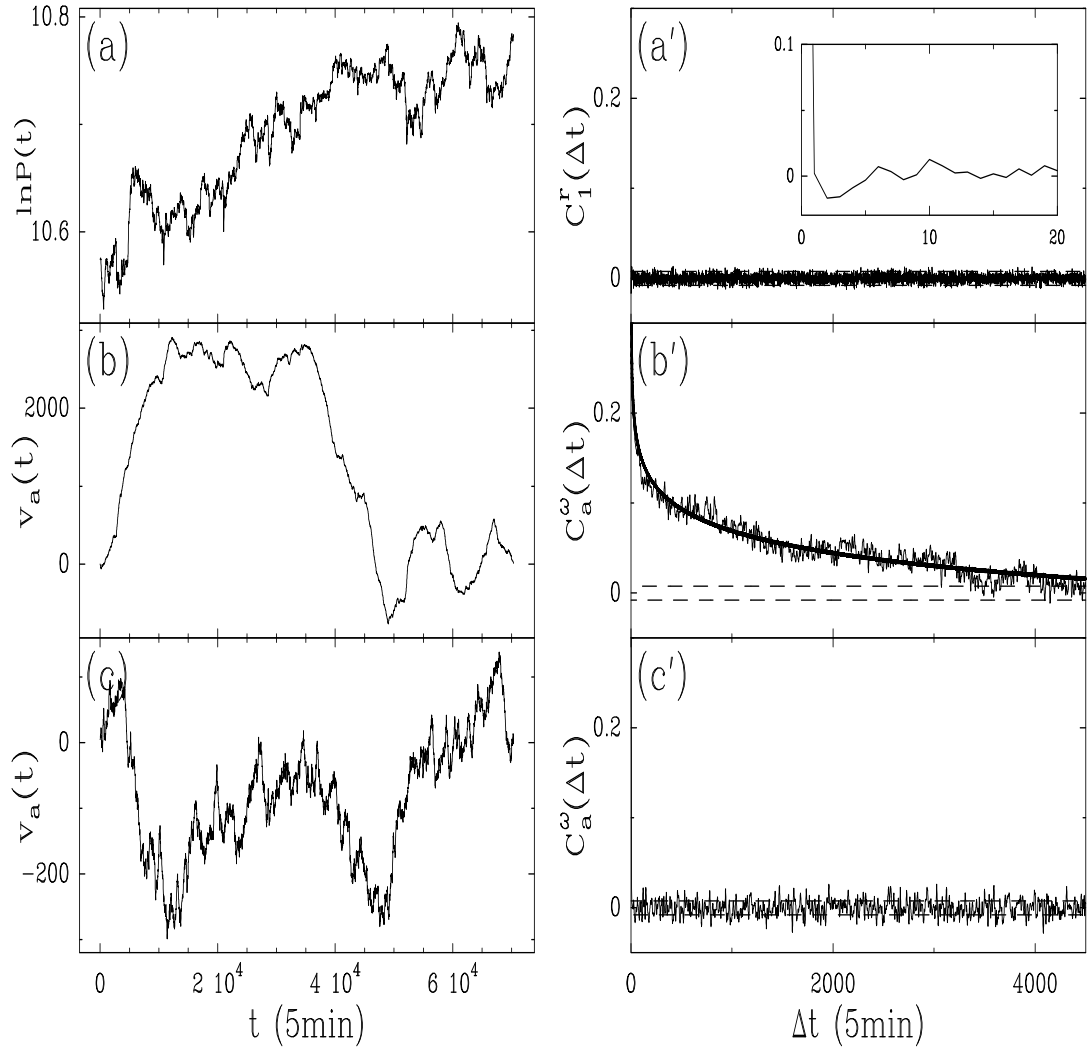,height=9cm,width=10cm,bbllx=140,bblly=330,bburx=490,bbury=530}
\caption{\protect\label{firstexstock}
(a) Time evolution of $\ln P(t)$, where $P(t)$ is the 
S\&P 500 US index, sampled with a time resolution $\delta t = 5$ min in 
the period October 1991-February 1995. The data have been preprocessed 
in order to remove ``parasitic'' daily oscillatory 
effects: if $m_i$ and $\sigma_i$
are respectively the mean and the r.m.s. of the signal within the i-th
5-min. interval of a day, the value of the signal $x(i)$ has been
replaced by $m+\sigma(x(i)-m_i)/\sigma_i$. (b) The corresponding
``centered log-volatility walk'', $v_a(t) = \sum_{i=0}^t \tilde{\omega}_a(i)$, as computed 
with the derivative of the Haar function as analyzing 
wavelet  for a scale $a = 4$ ($\simeq 20$ min). (c) $v_a(t)$ computed after having randomly 
shuffled the increments of the signal in (a). (a') The 5 min ($a = 1$) return
correlation
coefficient $C_1^r(\Delta t)$ versus $\Delta t$. (b')
The correlation coefficient $C_a^{\omega}(\Delta t)$ of the
log-volatility of the S\&P 500 at scale $a = 4$ ($\simeq$ 20 min); the
solid line corresponds to a fit of the  data using Eq.~(\ref{corr})
with $\lambda^2 \simeq 0.015$ and $T \simeq 3$ months. (c') same as in (b')
but for the randomly shuffled S\&P 500 signal. In (a'-c') the dashed
lines delimit the 95\% confidence interval. Taken from \cite{DSArnemuzy}.
}
\end{center}
\end{figure}

The attraction of physicists to finance and to the study of stock markets 
is grounded on several factors.
\begin{itemize}
\item Physics and finance are both fundamentally based on the theory of random walks
(and their generalizations to higher dimensions) 
and on the collective behavior of large numbers of correlated variables.
Finance thus offers another fascinating playground for the
application of concepts and methods developed in the Natural Sciences
which have traditionally focused their attention on a description
and understanding of the surrounding inanimate world at all possible scales.
\item Stock markets offer maybe one of the simplest real life experimental
system of co-evolving competing learning agents and can thus be thought of
as a proxy for studying biological evolution \cite{DSPalmer,DSCalda,DSFarmer,DSZhang}.
\item It is tempting to believe that the technical
abilities developed in the Physical Sciences could help to ``beat the market'':
predicting a complex time series like the market price evolution shown in 
figure \ref{firstexstock} is an exciting intellectual challenge as well as 
potentially rewarding financially.
\end{itemize}

In this short review, we present a series of models of stock markets that
each provide a particular window of understanding. The different models do not
play the same role. In its broadest sense, recall that a model (usually formulated using
the language of mathematics) is a mathematical representation 
of a condition, process,  concept, etc, in which the variables are defined
to represent inputs, outputs, and intrinsic states and
equations or inequalities are used to describe interactions of the 
variables and constraints on the problem. In theoretical physics, models
take a narrower meaning, such as in the Ising, Potts,..., percolation models.
In economy and finance, the term model is usually used in the broadest sense.
Here, we will use both types of models: 
The microscopic threshold models of the interplay between 
supply and demand discussed in section 2 and the percolation models 
of section 3 fall in the
second category. The 
cascade of correlations across scales \cite{DSArnemuzy} briefly
summarized in section 4 belong to this first
class of models. The log-periodic signatures preceeding crashes also discussed in 
section 4 relies
on both types of models. This diversity of models
reflects our burgeoning understanding of this field which has not yet fully
matured.

The first most striking observation of a stock market is that prices variations
seem to fluctuate randomly, leading to a price trajectory as a function of time
which looks superficially similar to a random walk with markovian increments,
as shown in the 
upper left panel (a) of figure \ref{firstexstock}. This view was first expoused
by Bachelier in his 1900 thesis \cite{DSBachelier} and later
formalized rigorously by Samuelson \cite{DSSamuelson}. This fundamental
thesis in finance is called the efficient market hypothesis and
states in a nutshell that price variations are essentially random as
a result of the incessant activity of traders who attempt
to profit from small price differences
(so-called arbitrage opportunities); the mechanism is that their investment strategies 
produce feedbacks on the prices that become random as a consequence. 
One important domain of research consists in determining the detailed
mechanisms by which this feedback operates dynamically and statistically.
Correlatively, the search of deviations from this efficient market state
may lead to significant understanding on the way the markets function.

At first glance, the concept that price variations are uncorrelated is confirmed
by looking at the two-point correlation function of the price increments.
For liquid markets such as the Standards and Poor's (SP\&500), it is
found significantly different from zero with statistical confidence
only for very short times of the order of a few minutes, 
as shown in panel (a') of figure \ref{firstexstock}.
On the other hand, the correlations of the amplitude of the variations,
called the volatility, are very long-ranged. This can be visualized
qualitatively by looking at panel (b) of figure \ref{firstexstock}
which constructs a random walk by successive addition of the logarithm
of a measure of the amplitude of the price variation obtained through
a wavelet transform (see \cite{DSArnemuzy} for details). One observes
long periods of persistences which can be compared with the random walk
in panel (c) obtained by first reshuffling the price variations
and performing the same analysis as for panel (b). Panel (b') and (c')
show the correlation functions of the volatilities corresponding
respectively to panels (b) and (c). For the real SP\&500 time series,
one observes an extremely slow approximately power law decay,
which provides a measure of the well-documented clustering or persistence of volatility.

In addition to the almost complete absence of correlation of price increments and
the long-range correlation of volatilities, the last striking stylized fact is the
``fat tail'' nature of the distribution of price variations or of returns.
The qualification ``fat tail'' is used to stress that the distributions of price
variations decay usually more slowly that a Gaussian, which is taken as the reference
that would be valid under the random walk hypothesis.
 Exponentially truncated L\'evy laws \cite{tMantegna,DScomment,dsBP}
with exponent around $\alpha \approx 1.5$ (see definition
(\ref{er})) for the 6-year period 1984-1989
and power laws with exponents $\alpha \approx 3$
\cite{dslux,dsgopi}, 
superposition of Gaussian motivated by an analogy with turbulence
\cite{DSGas,DSArnemuzy} or stretched exponentials \cite{dslaso} have been proposed
to describe the empirical distribution of price returns in organized markets.

In addition to these three well-documented stylized facts, 
many other studies have been performed to test for the possible existence of
dependence between successive price variations at many different time scales
that go beyond the method of correlation functions.
There is indeed increasing evidences that even the most competitive markets are not
completely free from correlations (i.e. are not
strickly ``efficient'') \cite{DSZhang} . In particular, a set of studies in
the academic finance literature have reported anomalous earnings which
support technical analysis strategies 
\cite{DSJoy,DSJagadeesh,DSLehmann,DSBrocketal}
(see \cite{DSMurphy} for a different view). A recent study of 60 technical
indicators on 878 stocks over a 12-year period \cite{DSBauerDah} finds that the
trading signals from technical indicators do on average contain information
that may be of value in trading, even if they generally underperform
(without taking due consideration to risk-adjustments of the returns) a
buy-and-hold strategy in a rising market by being relatively rarely invested.

In this review, we are going to present several models that propose
to explain some of these empirical observations. Microscopic models
of unbalance between supply and demand discussed in the next
section provide an understanding for 
the possible existence of different market phases, such as random, bubble-like
and cyclic as well as a simple mechanism for ``fat tails'' based on
multiplicative noise.
As we will summarize in section 3, the percolation models 
give probably the simplest
possible mechanisms for the observation of ``fat tails'' and long-range volatility 
clustering while being compatible with the absence of correlations of price
variations. They also teach us that finite size effects are probably important,
i.e. the number $N$ of traders that count on the market is perhaps no more than
a few hundreds to a few thousands. This is because most of the complexity 
observed in these and other similar models disappear when the limit of large $N$
is taken. If correct, this suggests that a suitable modeling of the stock market
belongs to the most difficult intermediate asymptotics between a few degrees of freedom and the
thermodynamic limit. As a consequence, it should take into account effects of 
discreteness. This might be one of the ingredients at the basis of the observation
of the remarkable log-periodic signatures preceeding large crashes discussed in section 4.
The accumulate
evidence now comprises more than 35 five crashes and 
to our knowledge, no major financial crash preceded by an extended bubble
has occurred in the past two decades without exhibiting log-periodic
signature. The exception is the 
East-European stock markets which seem to be following a completely different
logic than their larger Western counterparts and their indices does not
resemble those of the other markets. In particular, we find that they do not
follow neither power law accelerations nor log-periodic patterns though 
large crashes certainly occurs.

\section{Fluctuations of demand and supply in open markets}

\subsection{Optimization of supply faced to an uncertain demand}

Contrary to the common sense in economics, demand and supply do not balance in reality.
 You can find that all shelves are always full of commodities in any department 
 store in developed countries implying that supply is in excess. On the other hand, people are 
sometimes making queues in front of a popular bakery shop and fresh baked croissants 
 are sold out immediately, which clearly shows that demand is in excess. 
 Such excess-supply or excess-demand states can be shown to be maximal 
 profit strategies if we take into account the fluctuations of demand as follows.

Let us define the variables needed to describe the bakery's strategy: 
\begin{enumerate}
\item $x$, the selling price of a croissant; 
\item $y$, its production cost; 
\item $s$,  the production number of croissant per a day; 
\item $n$, the number of croissants requested by customers per day; 
\item $d$, the demand which is the averaged value of $n$. 
\end{enumerate}
We assume that $n$ fluctuates in the interval [$d-\delta, d+\delta$] 
uniformly for simplicity, and we also assume that the remainders will be 
thrown away at the end of each day. The problem is what is the best $s$ 
which maximizes the total profit. Obviously if $s=d+\delta$ then the 
bakery does not miss any customer's request and the gross sale is maximal. 
However, there is a possibility that it will have many unsold croissants 
when $n$ is small and in that case the production cost of the remainders 
may cause a big loss. At the other extreme, if  $s=d-\delta$, the bakery sells 
all its croissant and has no loss but on the other hand misses good selling 
opportunities. Therefore, there should be an optimal value of $s$ between these
two extremes that maximizes the expectation of the total profit.
 
Let us denote the expectation of the total profit by $L(s)$, then we have
 the following evaluation,
\begin{eqnarray}
L(s) & = & \langle x\: {\rm min}(n, s) - y s \rangle \nonumber \\
& = & (x - y)(d - \delta y/x) - \frac{x}{4 \delta} \{ s - d - \delta (1- 2y/x)\} ^2 \; \; \;
\end{eqnarray}
The maximal value of $L$ is given by the following value of $s$;
\begin{equation} 
s^*=d + \delta(1-2y/x) ~.     \label{eqpricbid}
\end{equation}
From this equation, it is clear that in the case where the sale price
is not very high, here $x<2y$, the optimal producing $s^*$ is smaller
that the average demand $d$. This corresponds to 
the excess-demand state which the popular bakery 
shop follows. On the other hand, if the sale price $x$ is higher than twice
the production price $y$, then the
best strategy is to keep the excess-supply state just like all department 
stores actually do ($s^*>d$). It should be noted that the balanced state of $s^*=d$ 
is the best strategy only when $x=2y$. This is the reason why almost all 
commodities in our daily life are out of the balance of demand and supply. 
The coefficient $2$ is of course modified if we assume a different probability 
density for the fluctuation of demand $n$. The key point in this discussion 
is the fluctuation of demand that is inevitable in any free economy society, 
and the best strategy taking such effect into account proves that the balance 
of demand and supply should almost always be broken to earn largest income
on average. 

A similar result is obtained if the bakery follows a different
strategy, i.e. strives to minimize its probability of loss: the probability
of losing is the same as the probability that 
$x~ {\rm min}(n, s) - y s$ be negative. In the interesting regime where $x>y$,
this probability is the same as the probability for the total sale $x n$ to be less
than the total production cost $y s$. This leads to a probability to lose equal to
\be
{\rm Proba}_{\rm loss} = {y \over 2 x \delta}~ [ s - s^{**}]~,
\ee
where
\be
s^{**} = (d-\delta){x \over y}~.
\ee
We see that the production $s^{**}$ that gives no loss with certainty is larger than the 
average demand $d$ only if the sale price $x$ is larger than ${d \over d-\delta} ~y$.

\subsection{Consequence for the bid-ask spread in liquid markets}

In an open market such as stock markets or foreign currency exchange markets, the 
situation is very different because there are speculative dealers who try to 
earn money by changing their position from a seller to a buyer or vise versa 
rather frequently. By this effect, the demand and supply can not be regarded 
as independent functions and furthermore we need to introduce a dynamic model to 
describe the pricing process correctly. Willing to buy from a market maker for instance,
you will buy a stock at the `ask' price $p_{\rm ask} \equiv x$ and resell it at the `bid' price 
$p_{\rm bid} \equiv y$. The spread $\delta p_{\rm spread} = p_{\rm ask} - p_{\rm bid}
= x-y$ is usually small. Indeed, 
the relevant situation for a liquid market is that the 
`ask' price $p_{\rm ask} \equiv x$ is only slightly larger than the 
`bid' price $p_{\rm bid} \equiv y$:
\be
x = y (1 + \epsilon)~,~~~~{\rm with}~ \epsilon << 1~.
\ee
Expanding (\ref{eqpricbid}) for small $\epsilon$ gives
\be
{s^* - (d-\delta) \over \delta} \approx {\delta p_{\rm spread} \over p_{\rm bid}}~.
\label{fjjkakla}
\ee
i.e. the relative over-supply with respect to the minimum possible value
$d-\delta$ is essentially equal to the relative spread. The implication of 
this result (\ref{fjjkakla}) is the following: reading (\ref{fjjkakla}) from
right to left, we find that a market maker will be tempted to
increase the spread between bid and ask if he has difficulty in getting rid
of excess inventory, but this will be a smaller effect, the larger are
the fluctuations of the demands, i.e. the possibility of selling in future
occasions.

\subsection{Microscopic model of market with threshold dynamics}

We assume that every dealer in an open market has two prices in mind, the selling and buying
 prices. For each dealer, the buying price is always lower than the selling 
 price, and the difference of these prices may represent his greediness.  
 A dealer's action is rather simple, namely, if the market price is higher 
 than the selling price in mind she will sell, and if the market price is 
 lower than the buying price in mind he will buy.  Let us assume the simplest 
 case that there are only two dealers, $A$ and $B$, and let their prices in 
 minds be, $p_{b}(A)$, $p_{s}(A)$, $p_{b}(B)$ and $p_{s}(B)$, where the 
 subscripts $b$ and $s$ represent the buying and selling prices and the 
 capital letters specify the dealers, $A$ and $B$. When these prices are 
 changed continuously, a trade occurs suddenly when either of the following 
 two conditions is realized \cite{tTakayasu92}:
\begin{equation}
p_{b}(A) \ge p_{s}(B) \; \; \; {\rm or} \; \; \; p_{b}(B) \ge p_{s}(A).
\label{eqn:ttrade}
\end{equation}
Note that the occurrence of a trade is characterized by a nonlinear function 
such as a step function.
 
As the greediness of the dealers always require $ p_{b}(A)< p_{s}(A)$  and 
 $p_{b}(B) < p_{s}(B)$, there is no possibility of realizing the two conditions 
 of Eq.(\ref{eqn:ttrade}) simultaneously, namely, the transaction is microscopically 
 one-sided or irreversible. After the trade, these dealers renew their prices in 
 their mind so that the trade condition does not hold any more.

Due to the nonlinear and irreversible nature of trades, dynamic models of dealers 
generally behave chaotically even if the dynamics is deterministic. There is a 
nonlinear effect that enhances any microscopic difference, but the estimated
 maximum Lyapunov exponent is 0 implying that the system is at the edge of 
 chaos \cite{tTakayasu92}. 

There are two extreme cases in this type of deterministic dealer models: one 
is the large asset limit and the other is the small asset limit. 
\begin{itemize}
\item In the case 
of large asset limit, dealers are assumed to have an infinite amount of asset and 
all dealers can keep their positions, namely, a buyer can be always a buyer 
and a seller can be always a seller. In this limit, it is shown that there is 
a kind of phase transition behavior between excess-demand and excess-supply 
states as a function of the number ratio of buyers to sellers.
In the excess-demand state, there are more buyers than sellers and
the prices fluctuate with a linear upgrade trend \cite{tHira}. In the
excess-supply phase, the situation is just opposite.  At the critical point,
that is realized when the numbers of buyers and sellers are the same,
there is no trend and the power spectrum of the price fluctuations
follow an inverse square law implying that the fluctuations are quite
similar to the Brownian motion.
\item In the small asset limit, each dealer changes position alternatively between 
a buyer and a seller, namely, after the dealer bought a stock, he tries to 
sell the stock. As all dealers change their positions alternatively, the 
numbers of demand and supply automatically balances and the system always 
shows critical behaviors, namely, the price fluctuations are similar to 
the Brownian motion even though the dynamics is deterministic \cite{tSato98}. 
This result indicates that the existence of speculative dealers who frequently 
change their positions is essential for the market to follow a random walk
scale-free behavior. Note that this kind of 
stationary self-organized criticality must be distinguished from 
the critical behavior describing large crashes as described in section 4.
The two phenomena are not mutually excluding as shown for instance in
ref.\cite{Huangetal}.
\end{itemize}

As dealers in any open market are sensitive to the market price changes, it is 
important to introduce a response effect in the dealer model to explain the 
fat-tail distribution of price changes as reported e.g. by Mantegna and 
Stanley \cite{tMantegna}. When dealers change their buying and selling 
prices in mind based on their own strategy independent of market price 
changes, the resulting price change distribution does not have long tails 
of power law. However, by adding the term that uniformly shifts all the 
dealers prices in mind proportional to the latest market price change, 
the distribution of market price changes become a power law in general \cite{tSato98}.           

\subsection{Derivation of Langevin market dynamics with multiplicative noise}

The reason for the fat-tail distributions can be theoretically explained by 
introducing a Langevin type stochastic equation with multiplicative noise:
\begin{equation}
\Delta P(t+\Delta t) = B(t)\Delta P(t) + F(t)
\label{eqn:teqlan}
\end{equation}
Here, $P(t)$ represents the market price at time step $t$ and $\Delta P(t) 
\equiv P(t) - P(t-\Delta t)$ is the price change where $\Delta t$ is 
the unit time interval. The effect of dealers' response on the market 
price change is given by $B(t)$, which is regarded as a random variable.
 The random additive term $F(t)$ is due to the chaotic behaviors inherent 
 in the dealer model. 
 
 In the low asset limit, it can be shown that 
the market price changes of 
 the deterministic dealer model nicely approximated 
 by the multiplicative stochastic process described by Eq.(\ref{eqn:teqlan}) \cite{tSato98}. 
 We now present a more direct deriavation of Eq.(\ref{eqn:teqlan}) by
considering the dealers' dynamics in a macroscopic way \cite{tTakayasu99}. 
Let $p_{b}(j,t)$ and $p_{s}(j,t)$ be the $j$-th dealer's buying and 
selling prices at time $t$, then the total balance of demand and supply 
in the market is described by the following function called the 
cumulative demand, $I(P,t)$;

\begin{equation}
I(P,t)=\sum_{j}{\Theta (p_{b}(j,t)-P)-\Theta(P-p_{s}(j,t))},
\end{equation}
where $\Theta (x)$ is the step function which is $0$ for $x<0$ and is $1$ 
for $x>0$. When $P$ is such that $I(P,t)>0$, the number of buyers is larger 
than that of sellers at the price. Therefore, the balanced price at time $t$, 
$P^*(t)$, is given by the equation, $I(P^*(t),t)=0$. It is a natural 
assumption for an open market that the price change in a unit time is 
proportional to $I(P(t),t)$ when the market price is $P(t)$; therefore, 
we have the following equation:

\begin{equation}
P(t+\Delta t) - P(t) \propto I(p(t),t).
\label{eqn:eqpi}
\end{equation}

As the buying and selling prices are not announced openly, no one knows the 
value of $P^*(t)$. Traders can only estimate it from the past market price data 
$ \{ P(t-\Delta t), P(t-2\Delta t),... \} $. Taking into account the effect 
that each dealer thinks in adifferent way, we can write down the time evolution 
equation of $P^*(t)$ as follows:

\begin{equation}
P^{*}(t + \Delta t) = P^{*}(t) + F(t) + W(P(t),P(t-\Delta t),...),
\label{eqn:eqpp}
\end{equation}
Here, $F(t)$ represents a random variable showing the statistical 
fluctuation of dealers' expectation, and $W(P(t), P(t-\Delta t),...)$ 
is the averaged dealers' response function. Considering the simplest 
non-trivial case, we have the following set of linear equations.

\begin{equation}
P(t+\Delta t) =P(t) + A(t) (P^{*}(t)-P(t))
\label{eqn:teqpa}
\end{equation}
\begin{equation}
P^{*}(t+\Delta t)=P^{*}(t)+F(t)+B(t)(P(t) - P(t-\Delta t)).
\label{eqn:teqpb}
\end{equation}
Here, $A(t)$ is given by the inverse of the slope of $I(P(t),t)$ at 
$P=P^*(t)$ which is proportional to the inverse of the price elasticity 
coefficient in economics, and $B(t)$ shows the dealers' mean response 
to the latest market price change, and both of these coefficients can 
be random variables.  If we can assume that $P(t)$ and $P^*(t)$ are 
always very close, the set of  Eqs.(\ref{eqn:teqpa}) and (\ref{eqn:teqpb}) 
become identical to Eq.(\ref{eqn:teqlan}).  Namely, if the market price 
always follows the motion of the balanced price and if the dealers' responses 
to the latest price change averaged over all the dealers fluctuates randomly 
for different time, then the market price fluctuation is well-approximated 
by the Langevin type equastion, Eq.(\ref{eqn:teqlan}).

It is well known that such a stochastic process (\ref{eqn:teqlan}) 
generally produces large 
fluctuations following power law distributions when $B(t)$ takes larger 
values than unity with finite probability 
\cite{DSChampenowne,DSKesten,DSHaan,DSCalan,DSSolomon,DSSorcont,tTakayasu97}.
A important condition to get a power law distribution is that the multiplicative noise
$B(t)$ must sometimes take values larger than one,
corresponding to intermittent amplifications. 
This is not enough: the presence of the additive term $F(t)$ (which can
be constant or stochastic) is needed to ensure a ``reinjection'' to finite values, 
susceptible to the intermittent amplifications. It was thus shown \cite{DSSorcont}
that (\ref{eqn:teqlan}) is only one among many convergent ($\langle \ln B(t) \rangle < 0$)
multiplicative processes with 
repulsion from the origin (due to the $F(t)$ term in (\ref{eqn:teqlan})) of the form
\begin{equation}
x(t+1) = e^{H(x(t), \{b(t), f(t),...\})} ~B(t) ~x(t)~,
\end{equation}
 such that $H \to 0$ for 
large $x(t)$ (leading to a pure multiplicative process for large $x(t)$)
 and $H \to \infty$ for $x(t) \to 0$ (repulsion from the origin). 
$H$ must obey some additional constraint
such a monotonicity which ensures that no measure is concentrated over a finite interval.
All these processes share the
same power law probability density function (pdf) 
\begin{equation}
P(x) = C x^{-1-\alpha}
\label{er}
\end{equation}
 for large $x$ with $\alpha$ solution of 
\begin{equation}
\langle B(t)^{\alpha} \rangle = 1~.
\end{equation}

The fundamental reason for the existence of the powerlaw pdf (\ref{er}) is that
 $\ln x(t)$ undergoes a random walk with
drift to the left and which is repelled from $-\infty$. A simple Boltzmann argument 
\cite{DSSorcont}
gives an exponential stationary
concentration profile, leading to the power law pdf in the $x(t)$ variable.

These results were proved for the process (\ref{eqn:teqlan}) by Kesten \cite{DSKesten}
using renewal theory and was then revisited by several authors
in the differing contexts of ARCH processes in econometry \cite{DSHaan} and 1D random-field Ising
models \cite{DSCalan} using Mellin transforms, and more recently using 
extremal properties of the $G -${\it
harmonic} functions on non-compact groups \cite{DSSolomon} and the Wiener-Hopf 
technique \cite{DSSorcont}. 

In the case 
that $B(t)$ depends on $\Delta P(t)$ especially when it does not take a 
large value if the magnitude of price change exceeds a threshold value, 
exponential cutoffs appear in the tails of distribution of price changes 
resulting in a more realistic distributions \cite{tSato98}. 

There are cases where the behaviors of the set of equations, 
Eqs.(\ref{eqn:teqpa}) and (\ref{eqn:teqpb}), deviate from that of 
Eq.(\ref{eqn:teqlan}). For example, in the special case that $B(t)$ is  
larger than $1$ and $A(t)$ is smaller than $1$ for a certain time interval 
then both $P(t)$ and $P^*(t)$ grow nearly exponentially and the difference 
of these values also grow exponentially. This case corresponds to the 
phenomenon called a bubble \cite{DSBlanchard}.
We can also find an oscillatory behavior 
of market price when $A(t)>1$. Namely, the set of price equations 
derived theoretically can show typical behaviors of second order 
difference equation for different parameter combinations, as also proposed
in Refs.\cite{DSBouchaudCont,DSFarmer}. Real data 
analysis based on this formulation is now under intensive study.

\section{Percolation Models}

\subsection{Basic percolation model of market price dynamics in $2$ to
infinite dimensions}

Besides the Levy-Levy-Solomon model \cite{dsLLS}, the Cont-Bouchaud model
\cite{dsCB} seems to be the one investigated by the largest number of 
different authors. It uses the well-known percolation model and applies
its cluster concept to groups of investors acting together. This percolation
model thus, similar to the random-field Ising markets \cite{dsiori}, applies
physics knowledge collected over decades, instead of inventing new models
for market fluctuations. 

In percolation theory \cite{dsbooks}, every site of a large lattice is
occupied randomly with probability $p$ and empty with probability $1-p$;
a {\it cluster} is a group of neighbouring occupied sites. For $p$ above some
percolation threshold $p_c$, an infinite cluster appears spanning the
lattice from one side to the opposite side. The average number
$n_s(p)$ of clusters containing $s$ sites each varies for large $s$ right
at the percolation threshold as a power law: 
\begin{equation}
  n_s \propto s^{-\tau}
\end{equation}
with an exponent $\tau$ increasing from about 2.05 in two dimensions to 5/2
in six and more dimensions. Close to $p_c$ a scaling law for large $s$ holds:
\begin{equation}
  n_s = s^{-\tau} f((p-p_c)s^\sigma)
\end{equation}
with $\sigma \simeq 0.5$.
For $p < p_c$, the cluster numbers decay asymptotically with a simple
exponential, while above $p_c$ they follow a stretched exponential with
log$(n_s) \propto -s^{1-1/d}$ in $d$ dimensions. 

Quite similar results are obtained if we switch from this site percolation
problem to bond percolation, where all sites are occupied but the bonds 
between nearest neighbours are occupied with probability $p$; then clusters
are groups of sites connected by occupied bonds.

For dimensionality $d > 6$, the critical exponents like $\tau$ are those
of the Bethe-lattice or mean-field approximation, invented by Flory in 1941,
where no closed loops are possible and for which analytic solutions are
possible: $\tau = 5/2, \, \sigma = 1/2, \, f = $ Gaussian.
In three dimensions, most percolation results are only estimated 
numerically. Infinite-range bond percolation is also called random graph 
theory; then every site can be connected with all other sites, each with 
probability $p$. This infinite-range bond percolation limit was selected
by Cont and Bouchaud \cite{dsCB} in order to give exact solutions,
while the later simulations concentrated on two- or three-dimensional site 
percolation, with nearest neighbours only forming the clusters.

For market applications, the occupied sites are identified with investors,
and the percolation clusters are groups of investors acting together. Thus
at each iteration, every cluster has three choices: all investors belonging
to the cluster buy (probability $a/2$); all of them sell (also probability
$a/2$); and none of them acts at this time step (probability $1-a$). Thus
the activity $a$ measures the time with which we identify one iteration:
if this time step is one second, $a$ will be very low since very few investors
act every second; if the time step is one year, $a$ will be closer to its 
maximum value 1/2. All investors trade the same amount, and have an infinite
supply of money and stocks to spend. Summation over all active clusters
gives the difference between supply and demand and drives the price $P(t)$:
\begin{equation}
R(t) = [P(t+1)-P(t)]/P(t) \propto \sum_{buy} n_ss - \sum_{sell} n_ss
\end{equation} 
In this way, the Cont-Bouchaud model has for a given lattice very few free 
parameters: the occupation probability $p$ and the activity $a$. Moreover,
algorithms to find the clusters in a randomly occupied lattice are known since
decades \cite{dsbooks}, and thus a computer simulation is quite simple if one
has already a working (Hoshen-Kopelman) algorithm to find clusters.

Without any simulation \cite{dsCB}, one can predict the results for very small
$a$. If for a lattice of $N = L^d$ sites, we have $a$ of order $1/N$, then 
typically no cluster, or only one, is active during one iteration. The price
change then is either zero or $\pm$ the size $s$ of the cluster. The 
distribution of absolute returns $|R|$ thus is identical to the cluster
size distribution $n_s$, apart from a large contribution at $R=0$. In 
particular, right at the percolation threshold $p=p_c$ we have a distribution
$\pi(|R|)$ of returns obeying a power law 
\begin{equation}
\pi(|R|) \propto 1/R^\tau
\end{equation}
for not too small $|R|$, similar to Mandelbrot's L\'evy-stable Pareto
distributions \cite{dsmandel}. The probability to have a jump of at least 
$|R|$ then
decays asymptotically as $1/|R|^\alpha$ with $\alpha = \tau -1$ between
1 and 3/2. The volatility or variance of the return distribution is thus
infinite at the percolation threshold, apart from finite-size and finite-time
corrections; the same holds for skewness and kurtosis.

For larger activities, but still $a \ll 1$, scaling holds \cite{dsSP}: 
if we normalize height and width of the return distribution to unity,
the curves for various activities $a$ at $p=p_c$ overlap, and thus still
give the above power law. This scaling is no longer valid for large $a 
\simeq 1/3$ where the curves become more like a Gaussian. 

This model thus reproduces some stylized facts of real markets, when inflation
effects are subtracted: i) The average
return $\langle R \rangle$ is zero. ii) There is no correlation between
two successive returns or two successive volatilities, since all active 
clusters decide randomly and without memory whether to buy or to sell, and 
since the occupied sites are distributed randomly. iii) At the percolation 
threshold, a simple asymptotic power law holds for small activities (short 
times) and becomes more Gaussian for large $a$ (long times).

This latter crossover to Gaussians, seen also in some analyses of real
markets \cite{dsgopi,dskull}, is not seen if we replace the percolation
model by a L\'evy walk for the price changes \cite{dsCS}, where the return
is a sum of steps distributed with the same power law exponent $\tau$ as
the above percolation clusters. In this simplification, the power law remains
valid also for large $a$ without a crossover to Gaussians. Note that in 
percolation, as opposed to L\'evy walks, the clusters are correlated by the
sum rule $\sum_s n_ss = pN$.

\subsection{Improvements of the percolation model}

\subsubsection{Clustering by diffusion}

Another advantage of this percolation model compared with L\'evy walks is 
volatility clustering. While $\langle R(t) R(t+1)\rangle$ in real markets
decays rapidly to zero (but see \cite{DSZhang} for different information), the 
autocorrelations of the absolute returns \cite{dsluxm} 
$\langle |R(t) R(t+1)|\rangle$ decay slowly as shown in panel (b') of figure
\ref{firstexstock} (see also e.g. Fig.2 of \cite{dsstanley}: 
a turbulent day on the stock market is often followed by another turbulent day,
though the sign of change for the next day is less predictable. We simulate
\cite{dsparis} this volatility clustering by letting the investors diffuse
slowly on the lattice; thus in the above picture, a small fraction of the 
investors move to another neighbourhood of the city where they get a different
advice from a different expert. Now the autocorrelation functions decay
smoothly, with unexplained size effects \cite{dsparis}.

\subsubsection{Feedback from the last price}

So far the model assumes the investors or their advisors to be complete 
idiots: They decide randomly whether to buy or to sell, without regard to
any economic facts. Such an assumption is acceptable for the author from
Cologne since the local stock market is in D\"usseldorf, not Cologne. However,
the discussions of log-periodic oscillations earlier in this review made clear
that not everything should be regarded as random. The simplest way to include
some economic reason is the assumption that prudent investors prefer to sell
if the price is high and to buy if it is low. Thus the probabilities to buy 
or to sell are no longer $a/2$ but are changed by an amount proportional to the
difference between the actual price and the initial price; the latter one is 
regarded as the fundamental or just price. Surprisingly, simulations 
\cite{dschang} show that the distribution $\pi(R)$ is barely changed; as
expected the price itself is now stabilized to values close to the fundamental
price. Little changes if we allow the fundamental price to undergo Gaussian
fluctuations as in \cite{dsluxm}. 

The distribution of the wealths of the investors can be investigated only if
one gives each investor a finite initial capital, and adds to it the profits
and subtracts the losses made by the random decisions to buy and sell. Bankrupt
investors are removed from the market. Simulations \cite{dslieb} give
reasonable return distributions, but in disagreement with reality \cite{dsLS}
no clear power laws with universal exponents.
 
\subsubsection{How to get the correct empirical exponent $\alpha \approx 3$?}
 
The above power law $\pi \propto 1/R^\tau$ with $2 < \tau \le 2.5$ may have been
sufficient some time ago \cite{dsmandel,dsmant} for which L\'evy stable 
distributions requiring $\tau = \alpha+1 < 3$ could be qualified,
but today's more accurate
statistics shows fat tails decaying faster with $\tau > 3$, 
though slower than a Gaussian. 
They may be such power laws multiplied with an
exponential function, also called truncated L\'evy distributions 
\cite{tMantegna,dsBP}, or stretched exponentials \cite{dslaso}, or most likely power laws
with an exponent near $\alpha = \tau-1 = 3$ \cite{dsgopi,dslux}.

Several ways were invented to correct this exponent and get $\alpha \simeq 3$.
One may work with $p$ slightly above $p_c$, where an effective power law with
$\alpha =3$ can be seen over many orders of magnitude \cite{dsparis}. (In this
case, as is traditional for percolation studies, one omits the contribution from
the infinite cluster.) Or one integrates over all $p$ between zero and the
percolation threshold, thus avoiding the question how investors work at $p=p_c$
without ever having read a percolation book \cite{dsbooks}; now $\alpha = \tau -
1+\sigma \simeq$ 1.5 to 2 \cite{dsSS}. Much better agreement with the desired 
$\alpha \simeq 3$ is obtained if we follow Zhang \cite{DSZhang} and take 
the price change $R$ not linear in the difference between supply and demand,
as assumed above, but proportional to the square root of this difference.
Then $\alpha = 2(\tau + \sigma -1)$ is about 2.9 in two dimensions, just as 
desired. Numerically \cite{dsSS}, this power law could be observed over five
orders of magnitude, similar to reality \cite{dsgopi}. 
Changing the activity $a$ proportionally to the last known price change breaks
the up-down symmetry for price movements; now sharp peaks in the price, with 
high activity, are followed by calmer periods with low prices and low activity
\cite{dsjan}.
 
\begin{figure}
\begin{center}
\epsfig{file=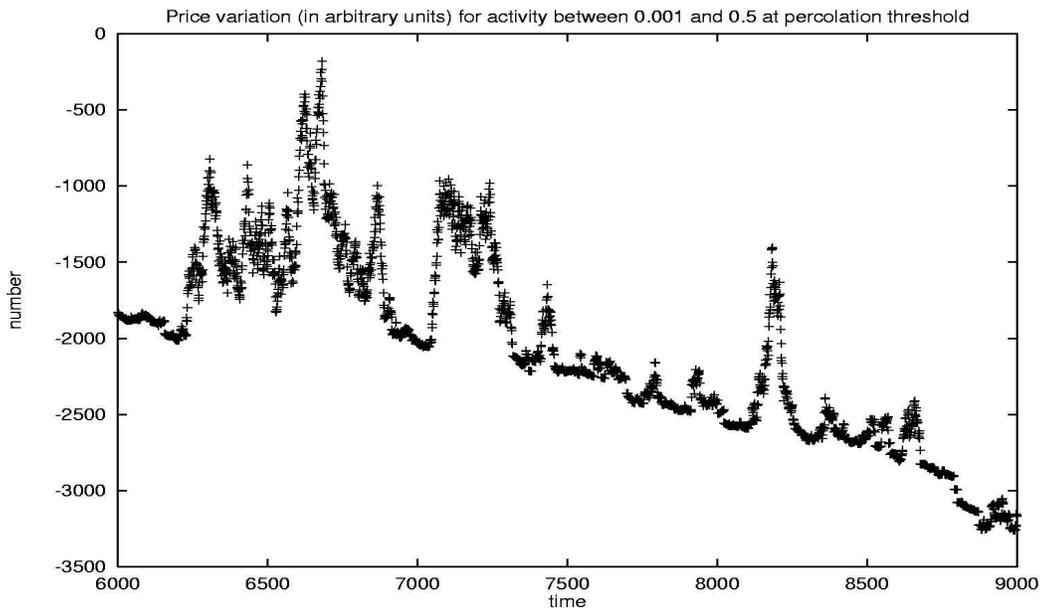, width=15cm, height=10cm}
\caption{\protect\label{stau1a} 
Single run of price versus time over 3,000 iterations, where the activity
is between $0.001$ and $0.5$. Such a simulation takes less than a minute on a
workstation. The units for the price change and the time are arbitrary. \cite{dsjan}
The choice of parameters exaggerates on purpose the
asymmetry between flat valleys and sharp peaks.
}
\end{center}
\end{figure}

Fig. \ref{stau1a} shows price change versus time, both
in arbitrary units, for 0.001 as the lower limit for the activity in the
model of \cite{dsjan}. 
Clearly, we see sharp peaks but not equally sharp holes (the downward trend
also indicates the survival probability of the first author if the Nikkei index 
fails to obey the prediction of Fig. \ref{predictNikkei} below). 
Fig. \ref{stau1b}  shows the desired slow
decay of the autocorrelation function for the volatility of this market model,
and for the same simulation Fig. \ref{stau1c}  gives the histogram of price changes.

\begin{figure}
\begin{center}
\epsfig{file=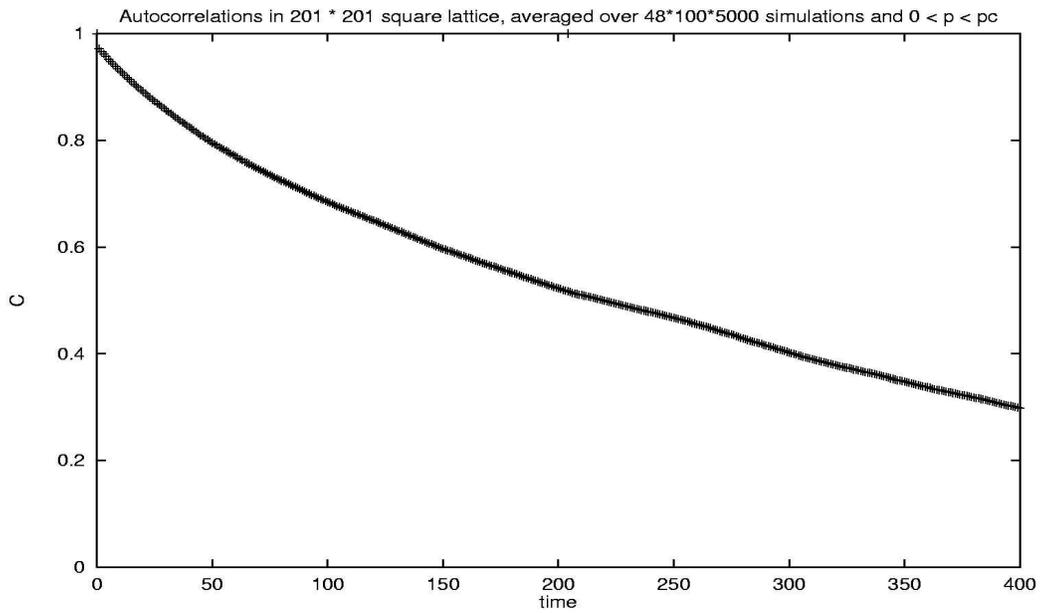,  width=15cm, height=10cm}
\caption{\protect\label{stau1b} 
Autocorrelations for the volatility, averaged over 4800 simulations similar
to figure \ref{stau1a}, requiring $10^2$ hours simulation time on a Cray-T3E. 
}
\end{center}
\end{figure}

\begin{figure}
\begin{center}
\epsfig{file=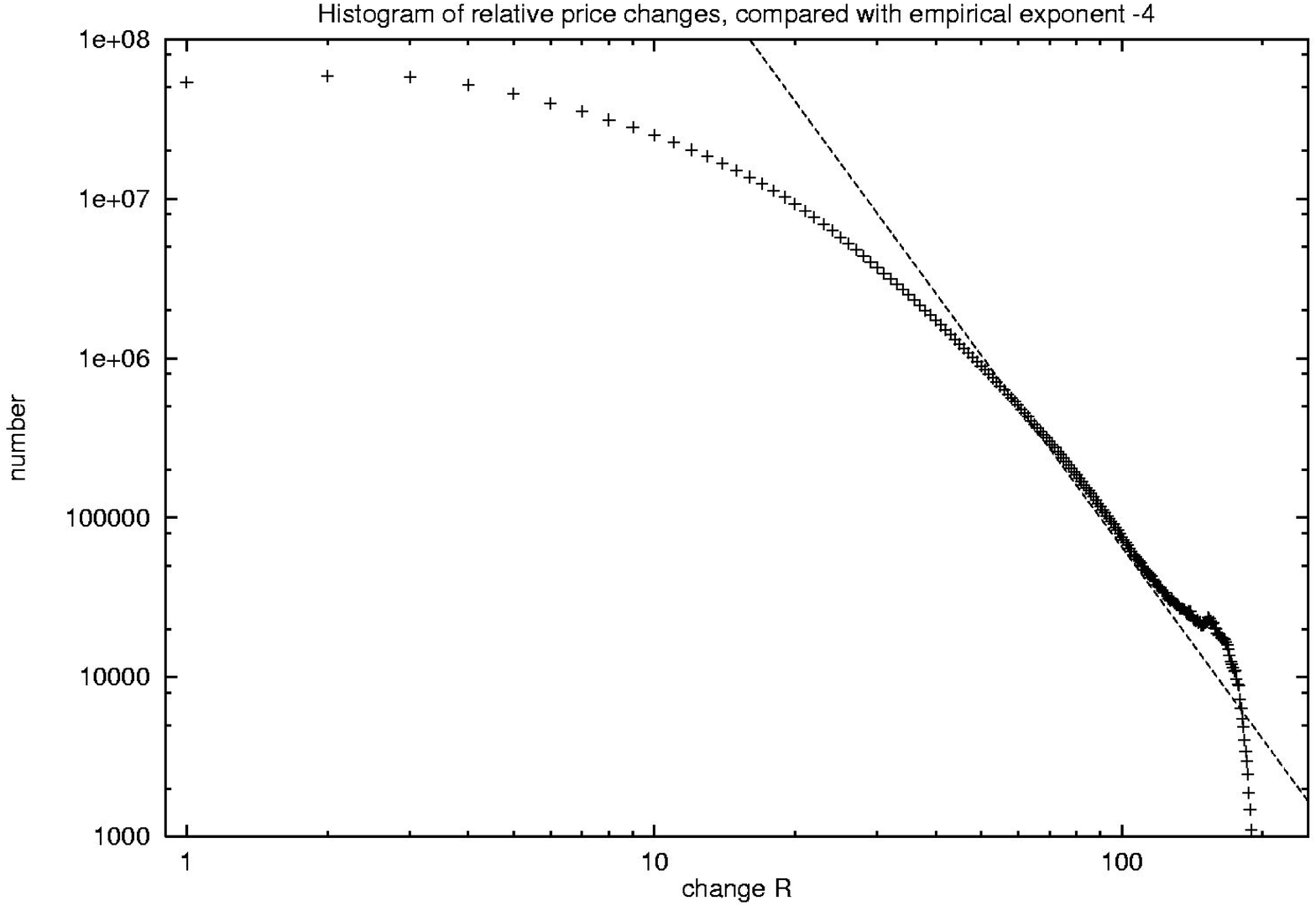,  width=15cm, height=10cm}
\caption{\protect\label{stau1c} 
Histogram for positive price changes on double-logarithmic scales. The
straight line has slope $-(1+\alpha) = -4$ \cite{dsjan}. Negative price changes 
show the same behaviour. 
}
\end{center}
\end{figure}

In the opposite direction, Focardi et al \cite{dsfocardi} assume the price
change to be {\it quadratic} in the difference between supply and demand
when the market gets into a crash. Using also other modifications of the
infinite-range Cont-Bouchaud model, their simulations show exponentially 
growing prices followed, at irregular time intervals, by rapid crashes.

\subsubsection{Log-periodicity and finite size effects}

None of these models has the ingredients which seem needed for log-periodic
oscillations before or after crashes, see  Sec.4.2. Percolation can give such oscillations
if we let particles diffuse on the occupied sites of the infinite cluster
for $p > p_c$, and if there is one preferred and fixed direction for this
diffusion (``bias'')\cite{dsbias}. Now the rms displacement of the diffusors
varies approximately as a power $t^k$ of time, and the effective exponent
$k(t)$ approaches unity with ocillations $\propto \sin(\lambda \ln(t))$. 
However, here the percolation clusters remain fixed while some additional
probing particle diffuses through the disordered medium; in the above
algorithm to produce volatility clustering, the investors themselves diffuse
and there is no additional probing particle. Thus these log-periodic 
oscillations of diffusive percolation have not yet been related by a simulation
to the Cont-Bouchaud percolation model for markets. 
 
The Cont-Bouchaud percolation model is particularly suited to look at effects
of finite lattice sizes, since size effects at such critical points have been
studied since decades. In most of the other microscopic models \cite{dswehia}, the
``thermodynamic limit'' $N \rightarrow \infty$ means that the fluctuations die
out or become nearly periodic. Real markets, according to these models, are 
dominated by the $10^2$ most important players and not by millions of
small investors. Also for the present Cont-Bouchaud model, the
behaviour becomes unrealistic (Gaussian $\pi(R)$) in this limit if $p < p_c$.
Right at $p=p_c$, however, the lattice is no longer self-averaging, and the
simulated return distributions keep the same shape for $N = 10^3$ to $10^6$.

Of course, the extreme tails are always dominated by size effects: No investor
can own more than 100 percent of the market, and no cluster can contain more 
than the $N = L^d$ lattice sites of the model. However, this trivial limit 
is relevant mainly above $p_c$; at the percolation threshold, the largest 
cluster is a fractal and contains on average $\propto L^D$ sites, where
the fractal dimension $D = d/(\tau-1)$ is smaller than $d$. Investigations of
the distribution of sizes for the largest critical cluster have only begun
\cite{dssen}.

\section{Critical crashes}

\subsection{Multiplicative cascades on the stock market}

The analogy between finance and hydrodynamic turbulence
developed by Ghashghaie {\em et al.} \cite{DSGas} 
implicitly assumes that 
price fluctuations can be described 
by a {\em multiplicative cascade}
along which the return $r$ at a given time scale $a < T$,
is given by:
\begin{equation}
   r_a(t) \equiv \ln P(t+a) - \ln P(t) = \sigma_{a}(t) u(t) \; ,
\label{proc}
\end{equation}
where $u(t)$ is some scale independent random variable, $T$
is some coarse ``integral'' time scale 
and $\sigma_a(t)$
is a positive quantity that can be
multiplicatively decomposed,
for any decreasing sequence of scales $\{a_i\}_{i=0,..,n}$ with
$a_0 = T$ and $a_n = a$, as \cite{DSGas}
\begin{equation} 
 \sigma_{a} = \prod_{i=0}^{n-1} W_{a_{i+1},a_{i}} \sigma_T \; .
\label{multi}
\end{equation}
Equation (\ref{proc}) together with (\ref{multi}) writes that the logarithm
of the price is a multiplicative process. But, this is different from
a standard multiplicative processes 
due to the tree-like structure of the correlations that are added by the
hierarchical construction of the multiplicands. We use
$\omega_a (t) \equiv \ln \sigma_a(t)$ 
as a natural variable.

If one supposes that $W_{a_{i+1},a_i}$  
depends only on the scale ratio $a_{i+1}/a_i$ and are i.i.d.
variables with log-normal distribution of mean $-H\ln 2$ 
and variance $\lambda^2 \ln 2$,
one can show \cite{DSArnemuzy} that the correlation function of the volatility
field $\omega_a (t)$
averaged over a period of length $T$ is given by
\begin{equation}
  \Gamma^{\omega}_a(\Delta t) = \lambda^2 \left(\log_2 {T\over \Delta t} -2 + 2 
{\Delta t\over T}\right) + \lambda_T^2 ~,  
\label{corr}
\end{equation}
for $a \leq \Delta t \leq T$ ($\left< . \right>$ means mathematical
expectation and $\lambda_T^2$ is the variance of $\omega_T$). 
For $\lambda^2 \simeq 0.015$ that can be 
obtained independently from the fit of the pdf's, 
Eq.~(\ref{corr}) provides 
a very good fit of the data (Fig 1(b')) 
for the slow decay of the correlation coefficient with only
one adjustable parameter $T \simeq 3 $ months.
Let us note that 
$C^{\omega}_a(\Delta t)$ can be equally well fitted by a power law $\Delta
t^{-\alpha}$
with
$\alpha \approx 0.2$. In view of the small value of $\alpha$, this is
undistinguishable from a logarithmic decay.
Moreover, Eq. (\ref{corr}) predicts that 
the correlation function $\Gamma^{\omega}_a(\Delta t)$ should 
not depend of the scale $a$ provided $\Delta t > a$ in agreement 
with data \cite{DSArnemuzy}.

Another very informative quantity is the cross-correlation function
of the volatility measured at different time scales:
\be
C^{\omega}_{a_1,a_2}(\Delta t) \equiv  \mbox{var}(\omega_{a_1})^{-1}
\mbox{var}(\omega_{a_2})^{-1}
\overline{\tilde{\omega}_{a_1}(t)\tilde{\omega}_{a_2}(t+\Delta t)}~.
\ee
It is found that $C^{\omega}_{a_1,a_2}(\Delta t) >
C^{\omega}_{a_1,a_2}(-\Delta t)$
if $a_1 > a_2$ and $\Delta t > 0$.
From the near-Gaussian properties of $\omega_a(t)$, the mean mutual
information of the variables $\omega_a(t+\Delta t)$ and
$\omega_{a+\Delta a}(t)$ reads\,: 
\begin{equation}
I_a(\Delta t,\Delta a) = -0.5 \log_2 \left( 1-
(C^{\omega}_{a,a+\Delta a}(\Delta t))^2 \right) \; .  \label{info}
\end{equation}
Since the process is causal, this quantity can be interpreted
as the information contained in $\omega_{a+\Delta a}(t)$
that propagates to $\omega_a(t+\Delta t)$.
The remarquable observation \cite{DSArnemuzy}  is the appearance of 
a non-symmetric propagation cone of information showing that the
volatility at large scales influences in the future 
the volatility at shorter scales. This clearly demonstrates of the pertinence of 
the notion of a cascade in market dynamics.

\subsection{Log-periodicity for ``foreshocks''}

A hierarchical cascade process as just described implies the existence of
a discrete scale invariance if the branching ratio and scale factor
along the tree are not fluctuating too much \cite{DSSornettereview}.
This possibility is actually born out by the data under the frame
of log-periodic oscillations.

\begin{figure}
\begin{center}
\epsfig{file=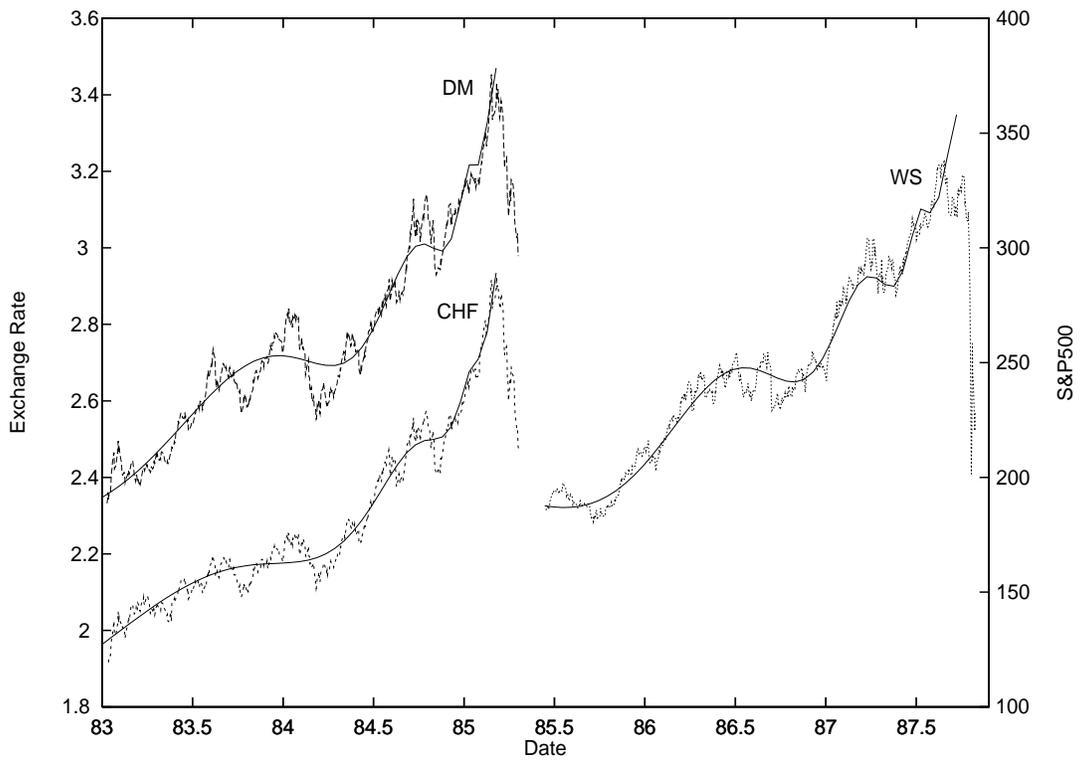,  width=15cm, height=10cm}
\caption{\protect\label{87} The S\&P 500 US index prior to the October
1987 crash on Wall Street and the US \$ against deutschmark (DEM) and 
Swiss franc (CHF) prior to the
collapse mid-85. The fit to the S\&P 500 is equation (\protect\ref{AE}) with
$p_c\approx  412  $, $ B \approx  -165 $, $
BC \approx   12.2 $, $  m \approx   0.33 $, $
t_c \approx   87.74  $, $ \Psi \approx   2.0 $, $ \lambda \approx   2.3$.
The fits to the DM and CHF currencies against the US dollar gives
$p_c\approx  3.88  $, $  B \approx  -1.2 $, $
BC  \approx   0.08 $, $  m \approx   0.28 $, $
t_c \approx   85.20  $, $ \Psi \approx   -1.2 $, $ \lambda \approx   2.8$ and
$p_c\approx  3.1  $, $ B \approx  -0.86 $, $
BC \approx   0.05 $, $ m \approx   0.36 $, $
t_c \approx   85.19  $, $ \Psi \approx   -0.59 $, $ \lambda \approx  3.3$,
respectively.  Reproduced from \cite{DSrisk}.
}
\end{center}
\end{figure}

As alluded to in the section on percolation models, log-periodicity refers is this context
to the accelerating oscillations that have been documented in stock market
prices prior and also sometimes following major crashes. The formula
typifying this behavior is the time-to-failure equation
\be
I(t) = p_c + B (t_c - t)^m \biggl[1 + C \cos\biggl(2 \pi
\frac{\log (t_c-t)}{\log \lambda} + \Psi \biggl) \biggl]~~~~,
\label{AE}
\ee
where $I$ is the price when the crash is a correction for a bubble
developing above some
fundamental value (it is the logarithm of the price if the crash
drop is proportional to the total price),
$t_c$ is the critical time at which the crash is the most probable,
$m$ is a critical exponent, and $\Psi$ is a phase in the
cosine that can be get rid of by a change of time units. 
$\lambda$ is the prefered scale factor of the accelerating oscillations
giving the ratio between the successive shrinking periods.
This expression (\ref{AE}) reflects a discrete scale invariance
of the price around the critical time, i.e. the price exhibits
self-similarity only under magnification around $t_c$ that are integer
powers of $\lambda$. Figure (\ref{87}) shows three cases
illustrating the behavior of market prices prior to large crashes \cite{DSrisk}.

Since our initial proposition of the existence of log-periodicity
preceding stock market crashes \cite{DSSJB96,DJSJ97,DSSJ98}, several works have extended the
empirical investigation
\cite{DSFF96,DS97crash,DSGluzman,DSVan1,DSVan2,DSmanisfesto,DSDrozdz}.

A recent compilation of many crashes \cite{DSrisk,DSjournrisk,DSIJTAF,DSEmpir}
provides increasing evidence of the relevance of log-periodicity and of the
application of the concept of criticality to financial crashes.
The events that have been found to qualify comprise
\begin{itemize}
\item the Oct.~1929, the Oct.~1987, the Hong-Kong Oct.~1987, 
the Aug. 1998 crashes, which are global market events,
\item the 1985 foreign exchange event on the US dollar,
\item the correction of the US dollar
against the Canadian dollar and the Japanese
Yen starting in Aug. 1998, 
\item the bubble on the Russian market
and its ensuing collapse in 1997-98,
\item twenty-two significant bubbles followed by large crashes or by severe
corrections in
the Argentinian, Brazilian, Chilean, Mexican, Peruvian, Venezuelan, Hong-Kong,
Indonesian, Korean, Malaysian, Philippine and Thai stock markets \cite{DSEmpir}.
\end{itemize}
In all these cases, it has been found that log-periodic power laws adequately
describe speculative bubbles on the western as well 
as on the emerging markets with very few exceptions.

The underlying mechanism which has been proposed \cite{DSrisk,DSjournrisk,DSIJTAF}
is that bubble develops by a slow build-up of long-range time
correlations reflecting those between traders leading eventually to a collapse of the
stock market in one critical instant. This build-up manifest itself as an
over-all power law acceleration in the price decorated by  ``log-periodic''
precursors. This mechanism can be analysed in an 
expectation model of bubbles and crashes which is essentially
controlled by a crash hazard rate becoming critical due to a collective
imitative/herding behavior of traders \cite{DSrisk,DSjournrisk,DSIJTAF}.
A remarkable {\it universality} is found for all events, with approximately the same value of
the fundamental scaling ratio $\lambda$ characterising the log-periodic signatures. 

To test for the statistical significance of these analyses,
 extensive statistical tests have been
performed \cite{DSIJTAF,DSThesis} to show that the reported ``log-periodic''
structures essentially never occurred in $\approx 10^5$ years
of synthetic trading following a ``classical'' time-series model,
the GARCH(1,1) model with student-t statistics (which has
a power law tail with exponent $\alpha = 4$), often used as a benchmark in
academic circles as well as by practitioners.
Thus, the null hypothesis that log-periodicity could result simply from random
fluctuations is strongly rejected.

\subsection{Logperiodicity for ``aftershocks''}

Log-periodic oscillations decorating an overall acceleration of the market
have their symmetric counterparts after crashes. It has been found
\cite{DSNikkeipaper} that
imitation between traders and their herding behaviour not
only lead to speculative bubbles with accelerating over-valuations of
financial markets possibly followed by crashes,
but also to ``anti-bubbles'' with decelerating market devaluations
following all-time highs. The mechanism underlying this scenario assumes that
the demand decreases
slowly with barriers that progressively quench in, leading to a
power law decay of the market price decorated by decelerating log-periodic
oscillations. This mechanism is actually very similar to that operating
in the random walk of a brownian particle diffusing in a random lattice
above percolation in a biased field \cite{dsbias}.

\begin{figure}
\begin{center}
\epsfig{file=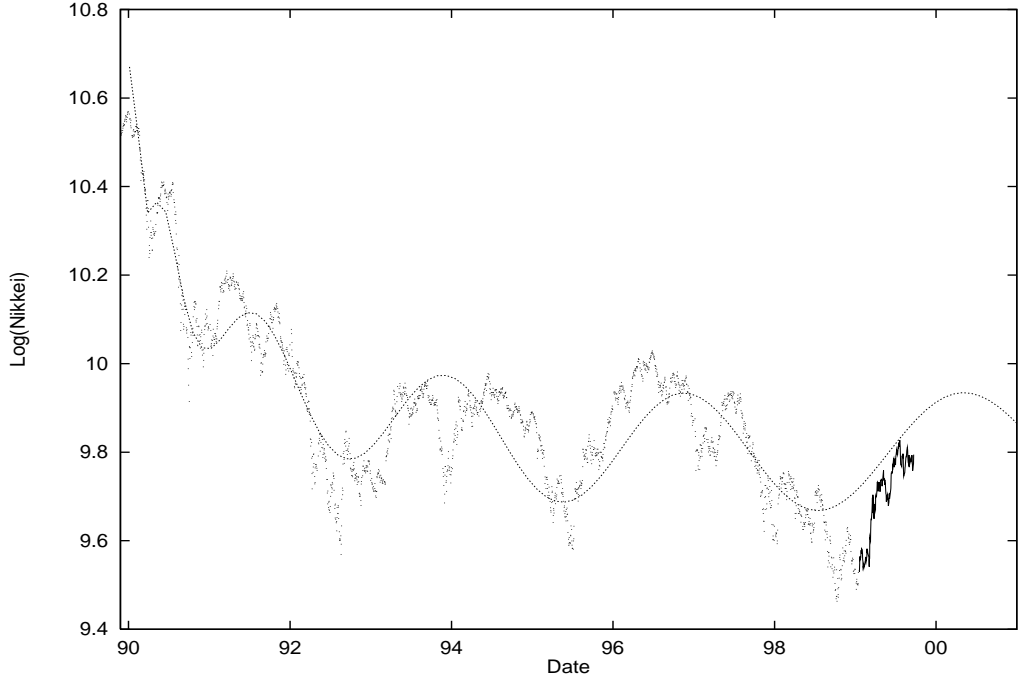,height=9cm,width=14cm}
\caption{\protect\label{predictNikkei} 
In \cite{DSNikkeipaper}, the Nikkei was fitted 
from 1 Jan 1990 to 31 Dec. 1998 with an extended log-periodic formula and
its extrapolation predicted that
the Japanese stock market should increase as the year 2000 was approached. In 
this figure, the value of the Nikkei is represented as the solid line 
after the last point used in
the analysis (31 Dec. 1998) until 21 Sep. 1999.
and can be compared with our prediction 
(the ticked line). The dots after Dec. 1989 until 31 Dec. 1998
represent the data used in the prediction.
This figure is as in \cite{DSNikkeipaper} except for the solid line starting
3rd Jan. 1999 which 
represents the realized Nikkei prices since the prediction was issued.}
\end{center}
\end{figure}

The strongest signal has been found on 
the Japanese Nikkei stock index from 1990 to
present
and on the Gold future prices after 1980, both after their all-time highs.
Figure \ref{predictNikkei} shows the Nikkei index representing the Japanese
market from the beginning of 1990 to present. The data
from 1 Jan 1990 to 31 Dec. 1998 has been fitted (the ticked line) by an 
extension of (\ref{AE}) using the next order terms in the expansion of 
a renormalization group equation \cite{DSNikkeipaper}. This fit has been
used to issue a forecast in early January 1999 for the recovery of the Nikkei in 1999
\cite{DSNikkeipaper}. The forecast, performed
at a time when the Nikkei was at its lowest, has correctly captured
the change of regime and the overall upward trend since the beginning of
this year. This prediction has first been released in january 1999 on the Los Alamos server
at http://xxx.lanl.gov/abs/cond-mat/9901268.  
The detailed publication for IJMPC \cite{DSNikkeipaper} was mentionned
already with its prediction in a wide-circulation journal which appeared in
May 1999 \cite{DSStaupredNi}.
One of the authors would not survive a Nikkei
drop since another author relied on the Nikkei prediction and invested in Japan.

A set of secondary western
stock market indices (London, Sydney, Auckland, Paris, Madrid, Milan, Zurich) 
as well as the Hong-Kong stock market have also been shown to exhibit well-correlated
log-periodic power law anti-bubbles over a period 6-15 months triggered by 
a rash of crises on emerging markets in the early 1994 \cite{DSEmpir}. As the US market 
declined by no more than $10\%$ during the beginning of that period and quickly
recovered, this suggests that these smaller stock western markets can ``phase 
lock'' (in a weak sense) not only because of the over-arching influence of Wall
Street but also independently of the current trends on Wall Street due to other
influences.

\section{Conclusion}

This review has attempted to present results that may advance our
understanding of the working of stock markets. First, we proved that 
the demand and supply should be deviating
from the balanced point in general cases when there are 
fluctuations in demand. In the case of an open market in which prices 
can change instantly following the unbalance of demand and supply, 
the speculative actions of dealers can be modeled numerically by models
with threshold dynamics.  The 
resulting market price fluctuations are characterized by a fat tail 
distribution when the dealers' response to latest price change is positive.
We have also shown that simple Langevin equation with
multiplicative noise account for the threshold-type dynamics of traders and
rationalize the ``fat tail'' nature of distribution of returns.
By solving the set of macroscopic market price equations, we have shown that 
there are three types in price changes: 1) stationary fluctuations, 2) bubble behavior, 
and 3) oscillatory phase. 
We believe that price fluctuations in open markets 
should be better understood by considering such dynamical effect that 
 has been neglected in ordinary approach of financial technology. 
 We have also presented models inspired by percolation that are 
 probably the simplest microscopic models capturing the effect of 
 imitation/clustering of traders in groups of various sizes.
 Improvements of the model provide reasonable agreement with the empirical value
 of the exponent $\alpha$ of the distribution of price variations. Clustering of volatility
 can also emerge rather naturally by feedback effect of the price on the 
 activity of the traders. The initial main weakness of the model, namely the fact
 that the connectivity had to be tuned to its critical value, has also been
 cured by allowing it to become a dynamical variable. The review ends up by
 summarizing the evidence for critical behavior associated with the formation of 
 speculative bubbles in large stock markets and their associated log-periodicity,
 corresponding to accelerated oscillations up to the time of crashes.
 Whether this will allow to prevent future crashes remains to be seen.
 
 The overall picture that emerges is quite interesting: a mixture of 
 more or less stationary self-similar statistical time series, maybe
 self-organized critical, with cascades of correlations accross time scales
 and once in a while a (truncated) divergence reflecting probably the
 crowd effect between traders culminating in a critical point with
 rather specific log-periodic signatures. According to the different models
 that we have presented, a crash has probably an endogenous origin and is
 constructed progressively by the market as a whole. 
In this sense, this could be termed a systemic instability. 
Further study might clarify what could be 
the regulations and informations that should be released to stabilize the market
and prevent these systemic instabilities.

\vskip 0.3cm

DS (fat, old, drunk, overcited)
thanks Naeem Jan for hospitality at St. Francis Xavier University where his
part of the review was written up. DS (thin, young, sober, undercited)
 is grateful to A. Johansen for preparing figures \ref{87} and
\ref{predictNikkei} and for a very stimulating and enjoyable collaboration over 
many years.

\end{document}